\documentclass[a4paper,aps,prd,10pt,preprintnumbers,showpacs,twocolumn,superscriptaddress,nofootinbib,amsmath,amssymb]{revtex4-1}
\usepackage[dvips]{graphics}
\usepackage[utf8]{inputenc}
\usepackage[T1]{fontenc}
\usepackage{cmap}

\def\imo{i}
\def\re#1{Re(#1)}
\def\im#1{Im(#1)}

\begin{document}
\title{Massive nonminimally coupled scalar field in Reissner-Nordström spacetime: Long-lived quasinormal modes and instability}
\author{R. A. Konoplya}\email{konoplya\_roma@yahoo.com}
\affiliation{Institute of Physics and Research Centre of Theoretical Physics and Astrophysics, Faculty of Philosophy and Science, Silesian University in Opava, CZ-746 01 Opava, Czech Republic}
\affiliation{Peoples Friendship University of Russia (RUDN University), 6 Miklukho-Maklaya Street, Moscow 117198, Russian Federation}
\author{Z. Stuchlík}\email{zdenek.stuchlik@fpf.slu.cz}
\affiliation{Institute of Physics and Research Centre of Theoretical Physics and Astrophysics, Faculty of Philosophy and Science, Silesian University in Opava, CZ-746 01 Opava, Czech Republic}
\author{A. Zhidenko}\email{olexandr.zhydenko@ufabc.edu.br}
\affiliation{Centro de Matemática, Computação e Cognição (CMCC), Universidade Federal do ABC (UFABC),\\ Rua Abolição, CEP: 09210-180, Santo André, São Paulo, Brazil}
\begin{abstract}
Here we show that the phenomenon of arbitrarily long-lived quasinormal modes (called \emph{quasiresonances}) of a massive scalar field in the vicinity of a black hole is not an artifact of the test field approximation, but takes place also when the (derivative) coupling of a scalar field with the Einstein tensor is taken into consideration. We observe that at large coupling and high multipole numbers, the growing modes appear in the spectrum, which are responsible for the eikonal instability of the field. For small coupling, when the configuration is stable, there appear the purely imaginary quasinormal modes which are nonperturbative in the coupling constant. At the sufficiently small coupling the nonminimal scalar field is stable and the asymptotic late-time tails are not affected by the coupling term. The accurate calculations of quasinormal frequencies for a massive scalar field with the derivative coupling in the Reissner-Nordström black-hole background are performed with the help of Frobenius method, time-domain integration and WKB expansion.
\end{abstract}
\maketitle

\section{Introduction}

The spectrum of quasinormal modes \cite{reviews} is an essential characteristic of a black hole. It governs the gravitational wave response to the external perturbations and, thereby, can be observed via gravitational interferometers \cite{TheLIGOScientific:2016src}. Quasinormal modes of a test scalar field have been very well studied by now for a great number of black-hole metrics, representing the Einstein theory and its alternatives. A number of papers in this area were devoted to quasinormal modes of test \emph{massive} fields. One of the motivations for such interest was the fact that at some fixed values of the mass of the field, special modes, called in \cite{Ohashi:2004wr} \emph{quasiresonances}, were observed. When approaching the above values of mass, the damping rate of these modes approaches zero \cite{Ohashi:2004wr}. This was shown for a massive test scalar field in the Schwarzschild \cite{Ohashi:2004wr,Konoplya:2004wg}, Kerr \cite{Konoplya:2006br} and Kerr-Newman \cite{Konoplya:2013rxa} backgrounds. Later, the same phenomenon of arbitrarily long-lived modes was observed for a massive vector \cite{Konoplya:2005hr} and Dirac \cite{Konoplya:2017tvu,Blazquez-Salcedo:2017bld} fields. Thus, apparently the effect does not depend on spin of the field. An effective quasiresonance phenomenon was noticed also for a massless scalar field around a black hole immersed in the asymptotically homogeneous magnetic field \cite{Wu:2015fwa}, because the magnetic field brings an effective massive term  \cite{Konoplya:2007yy}. In some spacetimes, such as the Schwarzschild-de Sitter one, the arbitrarily long-lived modes cannot exist, as it was shown analytically in \cite{Konoplya:2004wg}. Nevertheless, in the asymptotically de Sitter spacetimes a slower decay owing to the massive term was found \cite{Toshmatov:2017qrq}.

Arbitrarily long-lived quasinormal modes look like the standing waves and, thereby, describe rather an exotic situation. Unless the massive scalar hair exists compatibly with the black hole, the radiation should damp, and, in the end, the system must achieve the static state. The hint, that could probably resolve such an unnatural state, is that infinitely long-lived quasinormal modes are artifacts of the test field approximation. Indeed, any real physical interaction of a field in the vicinity of a black hole can never be represented \emph{exactly} as a pure test field in the black-hole background, even when the coupling terms might be neglected for practical reasons. Thus, the coupling terms representing for instance quantum corrections, coupling to other fields or classical backreaction of the field upon the black hole geometry would bring more realistic approximation to a real behavior of the field near the black hole. Therefore, we find it interesting to check whether there are quasiresonances in the spectrum of a massive scalar field with coupling corrections.

Reduction of the perturbation equations to a wavelike form in the presence of the coupling terms may be quite complicated problem which leads to the source term in the resultant master equations. This occurs, for example, when taking into consideration higher than linear terms of expansion in the perturbation theory. However, when one considers derivative couplings of a scalar field to the Einstein tensor, then the perturbation equations can be reduced to the traditional wavelike form with an effective potential and without the source term \cite{Chen:2010qf,Zhang:2018fxj,Yu:2018zqd}. In \cite{Yu:2018zqd} quasinormal modes of such a massive scalar field with derivative coupling were considered with the help of the third order WKB approach. The WKB formula, even when expanded to higher orders \cite{WKBorder}, cannot be used for accurate calculation of quasinormal modes of a massive scalar field, because the massive term adds an additional local minimum to the effective potential. Thus, there appear three turning points instead of the two for which the WKB formula of Schutz and Will and its higher order generalizations \cite{WKBorder} were developed. In addition, it is well known that the WKB method does not give reliable results for $n \geq \ell$, where $n$ is the overtone number and $\ell$ is the multipole number. Therefore, the dominant \emph{fundamental} mode $\ell=n=0$ was simply not calculated in \cite{Yu:2018zqd} even for the massless case.

Having all the above motivations in mind, our purpose here is twofold: first, to learn whether the arbitrarily long-lived quasinormal modes can survive when the coupling terms are taken into consideration and, second, to complement the calculations of \cite{Yu:2018zqd} by finding accurate quasinormal frequencies for the range of parameters which cannot be treated within the WKB approach \cite{WKBorder}. We shall use the three independent methods of calculations: time-domain integration \cite{Gundlach:1993tp}, the Frobenius method \cite{Leaver:1985ax} and the six order WKB formula \cite{WKBorder}. We will show that the arbitrarily long-lived quasinormal modes do exist when the derivative coupling is taken into account. It will also be shown that the late-time tails, which follow the period of the quasinormal ringing, are not affected by the coupling terms. We have found that the scalar field is unstable when the coupling constant is larger than some critical value, which agrees with \cite{Chen:2010qf}. Here we have shown that the instability is of the eikonal type, that is, it develops at high multipole numbers $\ell$. A similar instability was observed for the gravitational perturbations of black holes in the Einstein-Gauss-Bonnet and other higher curvature corrected theories \cite{Takahashi:2009xh,Konoplya:2017lhs,Grozdanov:2016fkt}. It is important, because the eikonal instability in some of the theories with higher curvature corrections is usually accompanied by the nonhyperbolicity of the corresponding perturbation equations \cite{Reall:2014pwa}, which means the breakdown of the whole regime of linear perturbations. In the formalism of linear perturbation this nonhyperbolicity  looks like the divergence of the wave function when summing over the modes with different (up to infinity) multipole numbers $\ell$ \cite{Cuyubamba:2016cug}.

The paper is organized as follows. Sec.~\ref{sec:basic} gives basic formulas on perturbation equations and the resultant wavelike equations with an effective potential. Sec.~\ref{sec:methods} briefly reviews the three methods used for calculations of quasinormal modes. Sec.~\ref{sec:data} is devoted to the discussion of the obtained numerical data, the observed instability and the deduction of an analytical formula for eikonal quasinormal modes. Finally, in the Conclusion, we briefly summarize the obtained results and mention open questions.

\section{Basic formulas}\label{sec:basic}

Here we shall consider a massive scalar field coupled to the Einstein tensor of the electrovacuum system. Thus, the background solution is given by the ordinary  Reissner-Nordström metric, while the massive scalar field is not described by the Klein-Gordon equation anymore. Instead, the equation of motion for a scalar field with the derivative coupling is used
\begin{equation}\label{1}
\square \Phi +\beta G_{\mu\nu}\nabla^{\mu}\nabla^{\nu}\Phi+\mu^2\Phi=0\,,
\end{equation}
where $\mu$ is the mass of the field and
$$G_{\mu\nu}=R_{\mu\nu}-\frac{1}{2}Rg_{\mu\nu}$$
is the Einstein tensor.
The nonminimal couplings of similar and more general forms have been recently considered in a number of papers, mostly in the cosmological context \cite{Sushkov:2009hk,Fontana:2018fof}. In \cite{Fontana:2018fof} quasinormal modes of a nonminimally coupled scalar field in the pure de Sitter spacetime have been computed and a dynamical instability has been found. In the Schwarzschild limit $Q=0$, the Einstein tensor vanishes, so that the configuration is reduced to a minimally coupled scalar field propagating in the Schwarzschild background.

The Reissner-Nordström black hole is described by the metric:
\begin{equation}
ds^2=-f(r)dt^2+\frac{dr^2}{f(r)}+r^2(d\theta^2+\sin^2\theta d\phi^2),
\end{equation}
where
$$f(r)=1-\frac{2M}{r}+\frac{Q^2}{r^2},$$
and $M$ and $Q$ are the mass and charge of the black hole.

The above equation of motion (\ref{1}) is reduced to the wavelike equation for the radial part $F(r)$ in the following way \cite{Yu:2018zqd}
\begin{equation}\label{eq:wavelike}
\left(\frac{d^2}{dr_*^2}+\omega^2-V\right)F(r)=0\;,
\end{equation}
where the tortoise coordinate is
$$dr_*=\frac{dr}{f(r)},$$
and the effective potential is given by
\begin{eqnarray}
V &=& \frac{{{r^2}-2Mr+{Q^2}}}{{{r^6}{{({r^4}+\beta {Q^2})}^2}}}\{{r^8}[{r^2}\ell(\ell+1)+{\mu^2}{r^4}+2Mr-2{Q^2}]\nonumber\\&&
+\beta{Q^2}{r^4}(6{r^2}+{\mu^2}{r^4}-12Mr+6{Q^2})\label{eq:pot}\\&&
+{\beta^2}{Q^4}[2{r^2}-\ell(\ell+1){r^2}-6Mr+4{Q^2}]\}\;.\nonumber
\end{eqnarray}
We shall parametrize the mass and charge of the black hole by the event horizon radius $r_+$ and the inner horizon $r_-$, so that
\begin{equation}
M=\frac{r_++r_-}{2}, \qquad Q^2=r_+r_-.
\end{equation}
Further we shall  use the dimensionless coupling
\begin{equation}
\alpha=\frac{\beta Q^2}{r_+^4}=\frac{\beta r_-}{r_+^3}>0.
\end{equation}

Let us notice that $\alpha$ is assumed to be non-negative, because of the singularity at $r=r_+\sqrt[4]{-\alpha}$ \cite{Yu:2018zqd}. Strictly speaking, we might also admit negative $\alpha>-1$, but such a coupling would imply existence of the minimal black-hole mass for which the singularity is still hidden by the horizon.

\section{The methods for finding quasinormal modes}\label{sec:methods}

By now there is extensive literature on the numerical and semianalytical methods for finding quasinormal modes of black holes. Therefore, we will only briefly review the methods used here, which are the Frobenius method, time-domain integration, and WKB expansion.

The quasinormal modes are solutions of the master wave equation (\ref{eq:wavelike}) which correspond to the purely incoming wave at the event horizon (because at the classical level a black-hole horizon does not reflect anything) and purely outgoing wave at spatial infinity. The latter condition is because the distant observer receives the incoming proper radiation of the black hole already when the source of perturbations stops acting and does not interfere with the observed signal.

\subsection{Frobenius series}

The Frobenius method is the most powerful approach to the calculation of the quasinormal modes, because it is based on the convergent procedure, so that, unlike the WKB formula, the frequencies can be obtained with any desired accuracy via Frobenius expansion. For the first time this approach was used for the search of quasinormal modes by Leaver  \cite{Leaver:1985ax}.

We represent the solution of (\ref{eq:wavelike}) as the Frobenius series near the event horizon
\begin{eqnarray}
\Phi(r)&=&e^{\imo\Omega r}(r-R)^{\lambda}\left(\frac{r-r_+}{r-R~}\right)^{\frac{-\imo\omega}{f'(r_+)}}\sum_{n=0}^\infty
a_n\left(\frac{r-r_+}{r-R~}\right)^n,\nonumber\\
\lambda&=&\imo(r_++r_-)(\Omega+\mu^2/2\Omega),\label{eq:Frobenius}
\end{eqnarray}
where
$$\Omega^2=\omega^2-\mu^2,$$
and the sign of $\Omega$ is chosen in order to have the outgoing wave (required by the quasinormal boundary condition) at spatial infinity; that is, for $\re\omega>0$ we choose $\re\Omega>0$ \cite{Konoplya:2004wg}.

The arbitrary parameter $R$, such that $r_-\leq R<r_+$, is chosen in order to satisfy
\begin{equation}\label{eq:inequality}
\left|\frac{r-r_+}{r-R~}\right|>1
\end{equation}
for all the singular points of the equation (\ref{eq:wavelike}), except the ones at the event horizon and spatial infinity \cite{Konoplya:2007jv}. Once $R$ is fixed in this way, we find the 15-term recurrence relation for the coefficients $a_n$ in (\ref{eq:Frobenius}).
Following \cite{Zhidenko:2006rs}, in order to find quasinormal modes, we reduce numerically the obtained recurrence to the three-term relation through Gaussian eliminations and use the continued fraction method \cite{Leaver:1985ax} together with the generalized Nollert improvement \cite{Nollert}.

\subsection{Time-domain profiles}

Although the Frobenius method allows one to find accurate values of quasinormal modes, one has to search in the frequency domain each mode
by minimizing the corresponding continued fraction. Alternatively one can use the time-domain integration of the perturbation equations, so that contribution of all the modes (at a given multipole number $\ell$) are taken into account within a single profile of ringing.
In order to produce the time-domain profiles, we integrate the wavelike equation (\ref{eq:wavelike}) rewritten in terms of the light-cone variables $u = t - r_*$ and $v = t + r_*$. The discretization scheme was described in detail in \cite{Gundlach:1993tp}:
\begin{eqnarray}\label{eq:scheme}
\Psi(N) &=& \Psi(W) + \Psi(E) -
\Psi(S)  \\
& &- \Delta^2\frac{V(W)\Psi(W) + V(E)\Psi(E)}{8} +
\mathcal{O}(\Delta^4)\ ,
\nonumber
\end{eqnarray}
where we have used the following definitions for the points: $N = (u + \Delta, v + \Delta)$, $W = (u + \Delta, v)$, $E = (u, v + \Delta)$ and $S = (u,v)$. The initial data are specified on the two null surfaces $u = u_{0}$ and $v = v_{0}$. The time-domain integration of a massive field does not allow one to extract the region of the quasinormal ringing precisely, so that there is some uncertainty in defining of the quasinormal frequencies. However, this method enables us to learn the behavior of the asymptotic tails at late times.

\subsection{WKB formula}

The approach is based on the WKB expansion of the wave function at both infinities (the event horizon and spatial infinity) which are matched with the Taylor expansion near the peak of the effective potential. The WKB approach in this form implies existence of the two turning points and monotonic decay of the effective potentials along both infinities
\begin{equation}\label{WKB}
	\frac{i Q_{0}}{\sqrt{2 Q_{0}''}} - \sum_{i=2}^{i=p}
		\Lambda_{i} = n+\frac{1}{2},\qquad n=0,1,2\ldots,
\end{equation}
where the correction terms $\Lambda_{i}$ were obtained in \cite{WKBorder} for different orders.  Here $Q_{0}^{i}$ means the $i$-th derivative of $Q = \omega^2 - V$ at its maximum with respect to the tortoise coordinate $r_*$, and $n$ labels the overtones. This approach can be applied to the modes with $\ell \geq n$ and to massless fields, as there are only two turning points in that case. For the massive field it is sufficiently accurate only when the mass term $\mu^2$ is not large \cite{Konoplya:2017tvu}, and the higher $\ell$, the larger $\mu^2$ are allowed. Although in the general case the WKB series converges only asymptotically, usually the difference between results obtained at higher and lower WKB orders gives an idea of how large the expected error of the approximation is.

\section{Quasinormal modes}\label{sec:data}

\subsection{Eikonal instability}

\begin{figure}
\resizebox{\linewidth}{!}{\includegraphics*{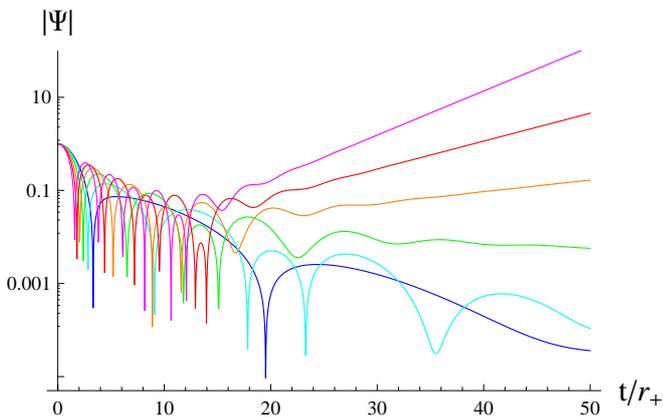}}
\caption{Time-domain profiles for the scalar field ($r_- = r_+/2$, $\alpha = 2$, $\mu = 0$) for stable multipole numbers $\ell=0$ (blue), $\ell=1$ (cyan), $\ell=2$ (green), and unstable $\ell=3$ (orange), $\ell=4$ (red), $\ell=5$ (magenta) .}\label{fig:instability}
\end{figure}

For large multipole number $l$ the effective potential (\ref{eq:pot}) takes the form
\begin{equation}\label{eq:potlargel}
V =\frac{f(r)}{r^2}\left(\ell+\frac{1}{2}\right)^2\frac{r^4-\beta Q^2}{r^4+\beta Q^2}+{\cal O}(1).
\end{equation}
From this one can see that for highly charged black hole, such that
\begin{equation}
r_+^4<\beta Q^2,
\end{equation}
the potential has an arbitrarily deep  negative gap at large $\ell$.
For such a sufficiently deep negative gap, a bound state with the negative energy is guaranteed, that means the instability. Indeed, in this situation for some finite value of $\ell$ a growing mode appears in the spectrum and dominates at late time (see Fig.~\ref{fig:instability}). For larger $\ell$, the instability grows faster, so that apparently $\ell\to\infty$ is the most unstable regime. From this it follows that one is not allowed to perform the multipole expansion of the wave function, so that the linear approximation should not be valid in the regime of instability. In other words, we expect that the hyperbolicity of the perturbation equations might be violated in this case.

This kind of instability at high multipole numbers $\ell$ was also observed in the spectrum of gravitational perturbations of a black hole with Gauss-Bonnet and other higher curvature terms (see, for example \cite{Takahashi:2009xh,Konoplya:2017lhs} and references therein). In a number of works  \cite{Grozdanov:2016fkt,Cuyubamba:2016cug} it was shown that the eikonal instability in the context of higher curvature corrected theories is triggered by the purely imaginary modes which are nonperturbative in the coupling constant in the sector of stability. When the coupling constant approaches zero, the imaginary part increases and goes to infinity \cite{Grozdanov:2016fkt,Cuyubamba:2016cug}. In other words the growing purely imaginary modes, responsible for the instability, do not go over into any known quasinormal modes at zero coupling, but instead, simply disappear from the spectrum.

\begin{table}
\begin{tabular}{|c|r|}
  \hline
  $\alpha$ & $\omega~r_+$ \\
  \hline
  $1.35$ &  $-0.037911 \imo$ \\
  $1.40$ &  $-0.013165 \imo$ \\
  $1.45$ &  $ 0.010513 \imo$ \\
  $1.50$ &  $ 0.033199 \imo$ \\
  $1.55$ &  $ 0.054965 \imo$ \\
  \hline
\end{tabular}
\caption{The purely imaginary mode which triggers the instability at $\alpha \approx 1.45$ ($r_-=r_+/2$) for $\ell =5$, $\mu =0$. The modes fit the linear law $\omega r_+= (0.4642 \alpha-0.6636) \imo $.}\label{tableIV}
\end{table}

\begin{figure}
\resizebox{\linewidth}{!}{\includegraphics*{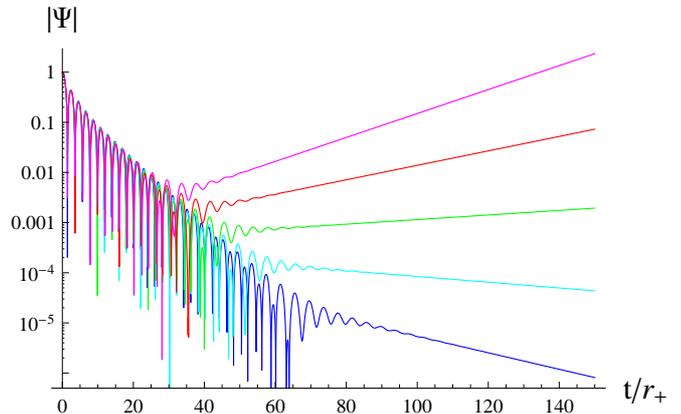}}
\caption{Time-domain profiles for the scalar field ($r_- = r_+/2$, $\ell = 5$, $\mu = 0$) for $\alpha=1.35$ (blue), $\alpha=1.40$ (cyan), $\alpha=1.45$ (green), $\alpha=1.50$ (red), $\alpha=1.55$ (magenta) .}\label{fig:threshold}
\end{figure}

Therefore, it would be interesting to know whether there are such kinds of purely imaginary, nonperturbative modes in the spectrum corresponding to the stable configuration of the scalar field. The thorough investigation of the spectrum via the time-domain integration on both sides from the threshold of instability, shows that, indeed, there is a set of purely imaginary modes whose damping rate increases when the coupling $\alpha$ is increased (see Table~\ref{tableIV}). The quasinormal frequencies as a function of $\alpha$ obey the linear law near the threshold of instability with very good accuracy. In order to distinguish the purely imaginary mode in a time-domain profile one should consider the later stages of the quasinormal ringing when the mode is not yet suppressed, but before the asymptotic tails become dominating (see Fig.~\ref{fig:threshold}). Thus, it would be numerically difficult to demonstrate that the purely imaginary frequencies which we found here indeed go to (minus) infinity, when $\alpha$ goes to zero. Nevertheless, in a similar fashion with the Gauss-Bonnet case, the tendency observed near the threshold of instability clearly indicates the nonperturbative character of these modes.

Having in mind that $\alpha=\beta Q^2/r_+^4$, we can see that
the potential (\ref{eq:pot}) is obviously positive definite for all values of $\alpha$ within the following range:
\begin{equation}
0\leq\alpha\leq1,
\end{equation}
which guarantees stability. On the contrary, the values of the coupling
\begin{equation}
\alpha > 1
\end{equation}
provide a negative gap, and, according to \cite{Takahashi:2009xh}, guarantee the abovementioned eikonal instability. In terms of quasinormal modes, the critical value of $\alpha$ corresponding to the threshold of instability is determined by the asymptotic behavior $\ell\to\infty$, while at each finite value of $\ell$ the growing mode appears at a larger than critical value of $\alpha$. The absence of convergence in $\ell$ indicates the lack of hyperbolicity of the master perturbation equation in the range of instability. For the massless scalar field the above instability was first observed in \cite{Chen:2010qf}, but no connection with the hyperbolicity was mentioned there. Here we have shown that the massive scalar field obeys the same threshold value of instability as the massless one, due to the subdominant contribution of any finite mass term in the eikonal regime.

\subsection{Quasinormal modes in the stable sector}

\begin{table}
\begin{tabular}{|l|c|}
  \hline
  Method & $\omega~r_+$   \\
  \hline
  Third order WKB & $0.573-0.165\imo$ \\
  Sixth order WKB & $0.575-0.173\imo$ \\
  Time-domain integration & $0.583-0.168\imo$ \\
  Frobenius & $ 0.5773-0.1646\imo$ \\
  \hline
\end{tabular}
\caption{Illustration of the concordance of data obtained by the three different methods: WKB, Frobenius and tim-domain integration. Here we have $r_- =0$, $\mu r_+ = 0.5$, $\alpha=1$ ($\ell=1$, $n=0$).}\label{TableII}
\end{table}

\begin{table*}
\begin{tabular}{|l|c|c|c|c|c|}
  \hline
  $\alpha$ & $\omega~r_+$ ($r_-=0$)& $\omega~r_+$  ($r_-=0.25r_+$) & $\omega~r_+$ ($r_-=0.50r_+$) & $\omega~r_+$ ($r_-=0.75r_+$) & $\omega~r_+$ ($r_-=0.95r_+$) \\
   \hline
  $0$   & $0.220910-0.209792\imo$ & $0.202534-0.168076\imo$ & $0.179447-0.132499\imo$ & $0.152670-0.109294\imo$ & $0.136881-0.098302\imo$\\
  $0.2$ & $0.206566-0.214425\imo$ & $0.192342-0.173834\imo$ & $0.173246-0.138765\imo$ & $0.150692-0.114506\imo$ & $0.136072-0.101361\imo$\\
  $0.4$ & $0.193183-0.218949\imo$ & $0.182967-0.179475\imo$ & $0.167776-0.144789\imo$ & $0.149029-0.119114\imo$ & $0.135278-0.104155\imo$\\
  $0.6$ & $0.180268-0.223410\imo$ & $0.174098-0.185167\imo$ & $0.162930-0.150744\imo$ & $0.147632-0.123248\imo$ & $0.134517-0.106731\imo$\\
  $0.8$ & $0.167401-0.227815\imo$ & $0.165494-0.191076\imo$ & $0.158681-0.156756\imo$ & $0.146460-0.126993\imo$ & $0.133799-0.109123\imo$\\
  $1.0$ & $0.154161-0.232111\imo$ & $0.156931-0.197403\imo$ & $0.155066-0.162909\imo$ & $0.145476-0.130410\imo$ & $0.133128-0.111359\imo$\\
  \hline
\end{tabular}
\caption{The fundamental quasinormal mode $\ell=n=0$ of the massless scalar field obtained by the Frobenius method for various  $r_-/r_+$ and $\alpha$.}\label{tableIII}
\end{table*}

The method which gives accurate values of the quasinormal modes is the Frobenius method. The other two methods, the time-domain integration and the WKB approach, are accurate only in some range of parameters. In table~\ref{TableII} one can see that in the range of parameters in which all the three  methods can be applied, the results are in a good concordance. The relatively small difference of time-domain integration and WKB results from the accurate Frobenius data is related to the fact that
(a) the WKB approach is more accurate for larger values of $\ell$ and smaller $\mu$,
and (b) within the time-domain integration for massive fields it is usually difficult to determine the region corresponding to the ring-down phase, because the asymptotic tail ``merges'' with the ringing phase.

\begin{table}
\begin{tabular}{|l|c|c|}
\hline
$\mu r_+$&$\omega~r_+$ ($r_-=0$)&$\omega~r_+$ ($r_-=r_+/8$)\\
\hline
$0$ &  $0.211801-0.212704\imo$&$0.204432-0.191646\imo$\\
$0.05$&$0.212085-0.211621\imo$&$0.204768-0.190559\imo$\\
$0.10$&$0.212900-0.208386\imo$&$0.205734-0.187315\imo$\\
$0.15$&$0.214145-0.203052\imo$&$0.207208-0.181975\imo$\\
$0.20$&$0.215681-0.195738\imo$&$0.209031-0.174685\imo$\\
$0.25$&$0.217388-0.186645\imo$&$0.211089-0.165685\imo$\\
$0.30$&$0.219240-0.176012\imo$&$0.213391-0.155229\imo$\\
$0.35$&$0.221302-0.164026\imo$&$0.216033-0.143476\imo$\\
$0.40$&$0.223664-0.150783\imo$&$0.219096-0.130480\imo$\\
$0.45$&$0.226381-0.136313\imo$&$0.222606-0.116252\imo$\\
$0.50$&$0.229462-0.120627\imo$&$0.226543-0.100796\imo$\\
$0.55$&$0.232891-0.103737\imo$&$0.230871-0.084129\imo$\\
$0.60$&$0.236636-0.085662\imo$&$0.235545-0.066283\imo$\\
$0.65$&$0.240662-0.066430\imo$&$0.240517-0.047287\imo$\\
$0.70$&$0.244928-0.046072\imo$&$0.245734-0.027171\imo$\\
$0.75$&$0.249398-0.024623\imo$&$0.251187-0.005977\imo$\\
\hline
\end{tabular}
\caption{The fundamental quasinormal mode $\ell=n=0$ obtained by the Frobenius method for $\alpha=1/8$ and various $\mu$.}\label{tableI}
\end{table}

From table~\ref{tableIII} we can learn that, once the coupling $\alpha$ is tuned on, the real oscillation frequency diminishes and the damping rate increases. The fundamental mode $\ell=n=0$ shown in tables~\ref{tableIII}~and~\ref{tableI} was not considered in \cite{Yu:2018zqd}, as the WKB formula gives a very large error in this regime, making the obtained results even inappropriate, for example, showing sometimes the growing (unstable) ``modes'' in the spectrum.

\begin{figure*}
\resizebox{\linewidth}{!}{\includegraphics*{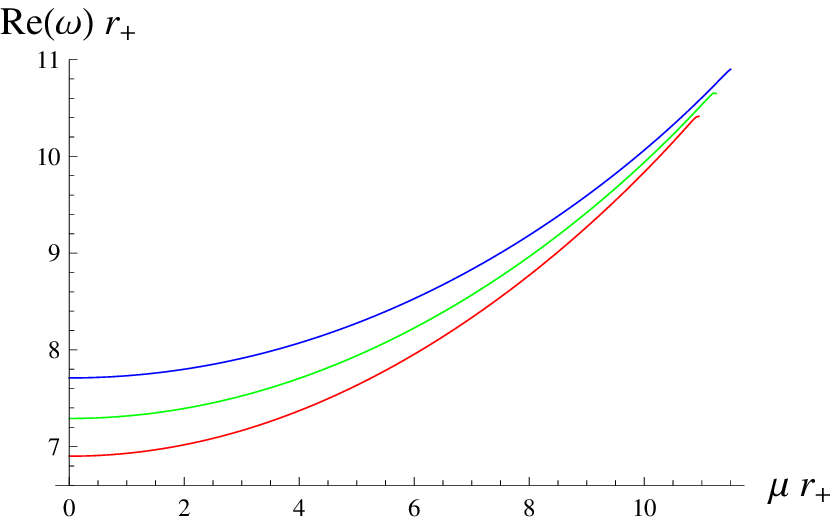}\includegraphics*{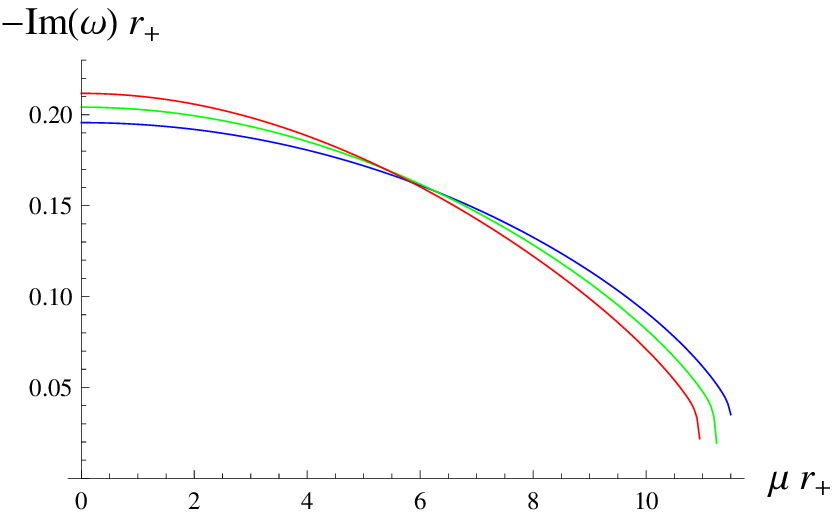}}
\caption{Evidence of quasiresonances in the eikonal regime: real (left panel) and imaginary (right panel) parts of the quasinormal modes for a weakly charged black hole $r_-= 0$, $\alpha = 1/8$ (blue), $\alpha =1/2$ (green), $\alpha =1$ (red) at a high multipole number $\ell = 20$.}\label{fig1}
\end{figure*}

In table~\ref{tableI} we can see that the damping rate of the fundamental quasinormal frequency $\ell=n=0$ diminishes up to very small values, when the mass $\mu$ is increased. Numerically, it is difficult to reach even smaller values of $\im\omega$, because the Frobenius method converges slower when approaching the regime of the nondamped modes -- quasiresonances. However, extrapolating the data from table~\ref{tableI} it is easy see that at some finite value of $\mu$ the damping rate vanishes. The WKB formula does not allow one to approach the regime of modes with very small damping rates as it was shown in \cite{Konoplya:2017tvu}. Nevertheless, in Fig.~\ref{fig1} one can see that at sufficiently large values of the multipole number $\ell$, when the WKB formula is accurate enough, the corresponding modes tend to those with zero damping rate. A similar picture can be obtained as a result of time-domain integration, though, as it was mentioned before, the ringing period will be hardly distinguishable from the asymptotic tails. Thus, we have clear numerical evidence that a massive scalar field with the derivative coupling to the Einstein tensor has quasiresonances in its spectrum. Here we have also showed numerical data for the case $r_{-}/r_{+} =0$, but $\alpha \neq 0$, representing the black hole whose charge is negligible (which is astrophysically relevant), but, at the same time, the coupling with the Einstein tensor is not small.

\begin{figure}
\resizebox{\linewidth}{!}{\includegraphics*{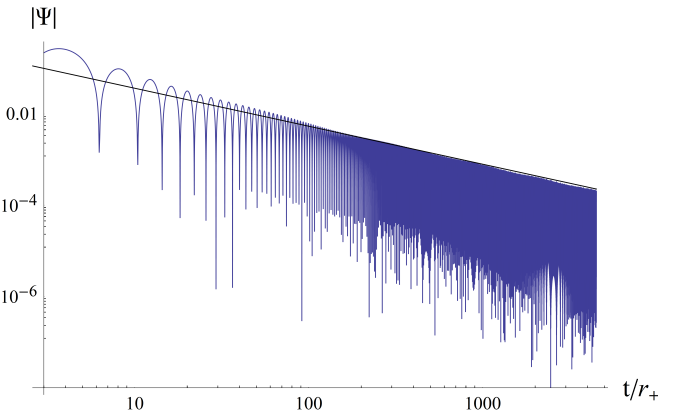}}
\caption{Quasinormal ringing and asymptotic tails for a weakly charged black hole $r_-/r_+ = 0$, $\alpha = 1$, $\mu r_+ = 1$.}\label{fig2}
\end{figure}

At asymptotically late times, the following power-law behavior is observed in Fig.~\ref{fig2}:
\begin{equation}
\Psi (t) \propto t^{-\frac{5}{6}} \sin (\mu \cdot t),
\end{equation}
where $\omega(t)$ depends weakly on time. The same law was observed for a test massive scalar field in the Schwarzschild \cite{Koyama:2001ee}, Reissner-Nordström \cite{Koyama:2000hj} and Kerr \cite{Burko:2004jn} backgrounds.

In \cite{Cardoso:2008bp} it was argued the existence of an interesting correspondence, stating that quasinormal modes which any stationary, spherically symmetric and asymptotically flat black hole emits in the eikonal regime are determined by the parameters of the circular null geodesic: the real and imaginary parts of the quasinormal mode are multiples of the frequency and instability timescale of the circular null geodesics respectively. In \cite{Konoplya:2017wot}  it was shown that the correspondence indeed exists for a number of cases; it is guaranteed for test fields in the vicinity of a black hole, but not for the gravitational field itself. Here we shall learn whether the correspondence works for a nonminimal scalar field. For this purpose we shall derive the analytical formula for quasinormal modes in the eikonal ($\ell\to\infty $) regime.

In the eikonal regime ($\ell\to\infty$) the analytical formula for quasinormal modes can be deduced by the first order WKB formula
\begin{equation}\label{eq:WKB1}
	\frac{i Q_{0}}{\sqrt{2 Q_{0}''}} = n+\frac{1}{2},
\end{equation}
applied to the potential (\ref{eq:potlargel}).
For small values of $\alpha$ we find that the maximum of the potential is located at
\begin{equation}\nonumber
r=r_0\left(1+\dfrac{4\alpha r_+^4(r_0-r_+)^2}{r_0^4(3r_0^2-8r_0r_++6r_+^2)}\right)+{\cal O}\left(\alpha^2,\ell^{-2}\right),
\end{equation}
where $r_0$ is the point of maximum for $\alpha=0$,
\begin{eqnarray}\label{maximum}
r_0&=&\frac{3(r_++r_-)+\sqrt{9(r_+-r_-)^2+4r_+r_-}}{4}\\
&=&\frac{3M+\sqrt{9M^2-8Q^2}}{2}.\nonumber
\end{eqnarray}

Then, using (\ref{maximum}) in (\ref{eq:WKB1}) and expanding the result for $\omega$ in terms of $1/\ell$, we obtain
\begin{eqnarray}\nonumber
&&\omega r_+=\frac{r_0-r_+}{2r_0}\sqrt{\frac{6r_0-r_+-9r_-}{r_0}}\Biggr[\left(\ell+\frac{1}{2}\right)\left(1-\frac{\alpha r_+^4}{r_0^4}\right)\\
&&-\imo\left(n+\frac{1}{2}\right)\left(1+\alpha K\right)\sqrt{2-\frac{3r_++3r_-}{2r_0}}\Biggr]+{\cal O}\left(\alpha^2,l^{-1}\right),\nonumber
\end{eqnarray}
where
$$K=3\left(1-\frac{r_+}{r_0}\right)^2\left(\frac{1}{2}-\frac{r_+^2}{r_0^2}\right)\left(\frac{6r_0-r_+-9r_-}{4r_0-3r_+-3r_-}\right)^2>0.$$
Let us notice that this expression can only be used to calculate the damped modes of the spectrum far from the threshold of instability, as the WKB formula (\ref{eq:WKB1}) cannot be applied to finding of the frequencies for which $Re (\omega) \ll Im (\omega)$.

Since the null geodesic of the Reissner-Nordström black hole remains unaffected, we find that the frequency and instability timescale differ, respectively, by
\begin{equation}\label{eq:delta}
\dfrac{\Delta\Omega_c}{\Omega_c}=\dfrac{\alpha r_+^4}{r_0^4}+{\cal O}(\alpha^2), \qquad \dfrac{\Delta\tau_c}{\tau_c} = \alpha K +{\cal O}(\alpha^2),
\end{equation}
from the real and imaginary part of the quasinormal modes in the eikonal limit.

In the limit of the weakly charged black hole ($r_-\to0$) (\ref{eq:delta}) reads
$$
\dfrac{\Delta\Omega_c}{\Omega_c}=\left(\dfrac{2}{3}\right)^4\alpha+{\cal O}(\alpha^2), \qquad \dfrac{\Delta\tau_c}{\tau_c} = \left(\dfrac{2}{3}\right)^5\alpha +{\cal O}(\alpha^2).
$$

Thus, the correspondence is not guaranteed for nonminimally coupled fields as well, which is in concordance with \cite{Konoplya:2017wot,Chaverra:2016ttw}.

\section{Conclusions}

Although the diminishing of the damping rate of quasinormal modes owing to the nonzero mass of the field has been known for a long time \cite{Konoplya:2002wt} (see also more recent papers \cite{Hod:2016jqt}), the complete vanishing of the damping rate observed first in \cite{Ohashi:2004wr} looks already unnatural, so that studying more realistic configurations, which include coupling terms, could promise a resolution of this ``paradox'' of quasiresonances. Here we have studied such a coupling of the scalar field with the Einstein tensor and found that the arbitrarily long-lived quasinormal modes remain even in this case. Thus, apparently the effect of quasiresonances is not an artifact of the test field approximation, but valid for more realistic configurations, admitting the coupling terms. In this connection it would be interesting to consider other types of couplings of a scalar field to gravitational and other fields, which could model the nonminimal interaction between the gravitational background and the scalar field, and, first of all, to study perturbations of higher than linear order.

Here we have also complemented calculations of quasinormal modes done in \cite{Zhang:2018fxj} by finding fundamental (and therefore dominating at late times) quasinormal modes $\ell=n=0$ which were omitted in \cite{Zhang:2018fxj}, because the WKB  formula used there does not work for this case. The time-domain integration which we used allowed us to show that the asymptotic tails are the same at the nonzero derivative coupling as they are for the test massive scalar field.

We have shown that for the coupling $\alpha > 1$ the scalar field is unstable. The instability occurs at large multipole numbers $\ell$ and it is similar to the eikonal instability of Gauss-Bonnet black holes in a number of aspects. The new branch of modes which are purely imaginary and nonperturbative in the coupling $\alpha$ has been found. These modes trigger the instability. As higher values of $\ell$ correspond to the most unstable part of the spectrum, it is evident that there is no convergence in $\ell$ and in the range of eikonal instability, the hyperbolicity is also lacking. This is a breakdown of the regime of linear perturbations, which apparently signifies that the configuration at large coupling $\alpha > 1$  cannot be considered self-consistently.

It is interesting to notice that, an instability of the Reissner-Nordström black hole has been recently found also for the scalar field coupled to the Maxwell field \cite{Myung:2018vug} at sufficiently large coupling. However, that instability develops at the lowest multipole number and does not lead to the violation of hyperbolicity of the perturbation equations.

Finally, in the stable regime, we have obtained the eikonal analytical formula for quasinormal modes and showed that for the nonminimal coupling it does not coincide with the parameters of the null geodesics as expected in \cite{Cardoso:2008bp}.

The eikonal instability apparently should appear for other spherically symmetric backgrounds and our paper could be extended in this direction. It would also be interesting to generalize the present work to the Kerr-Newman spacetime, thus taking into consideration the effect of rotation of a black hole.

\acknowledgments{
A.~Z.~was supported by Conselho Nacional de Desenvolvimento Científico e Tecnológico (CNPq), Brazil. R.~K.~acknowledges support of the International Mobility Project CZ.02.2.69\slash0.0\slash0.0\slash16\_027\slash0008521 and hospitality of the Silesian University in Opava. Z.~S.~acknowledges the Albert Einstein Centre for Gravitation and Astrophysics supported under the Czech Science Foundation (Grant No. 14-37086G).}

\end{document}